\documentclass[
    ,final            
  ]
  {aipproc}

\layoutstyle{6x9}
\usepackage{wrapfig,rotating}
\usepackage{amssymb,amsmath,array}
\usepackage{mathrsfs}

\newcommand{\bea}{\begin{eqnarray}}
\newcommand{\eea}{\end{eqnarray}}

\begin{document}

\title{Double-Spin Asymmetry $A_{LT}$ 
for
Polarized Drell-Yan Process in $p\bar{p}$ Collisions: Wandzura-Wilczek Contribution}


\author{Yuji Koike}{
  address={Department of Physics, Niigata University,
Ikarashi, Niigata 950-2181, Japan}
}

\author{Kazuhiro Tanaka}{
  address={Department of Physics, Juntendo University, Inba, Chiba 270-1695, Japan}
}

\author{Shinsuke Yoshida}{
  address={Department of Physics, Niigata University,
Ikarashi, Niigata 950-2181, Japan}
}

\begin{abstract}
The longitudinal-transverse spin asymmetry $A_{LT}$ in the polarized
Drell-Yan process depends on twist-3 quark distributions of nucleon. 
In addition to the contributions associated with the 
twist-3 operators,
these distributions contain the ``Wandzura-Wilczek (WW)'' part,
which is determined by a certain integral of twist-2 distributions.
The recently obtained empirical information on the transversity
allows a realistic estimate of the WW contribution to $A_{LT}$ 
for the polarized $p\bar{p}$ collisions.
Our results indicate 
that rather large $A_{LT}$ ($\sim 10$\%) 
can be observed in the proposed spin experiments at GSI, 
with novel pattern as a function of dilepton mass compared with $A_{TT}$ and $A_{LL}$.
\end{abstract}

\maketitle

The proposed polarization experiments 
with antiprotons at GSI provide promising opportunities to probe
chiral-odd 
spin-dependent parton distributions $h_1$, $h_L$~\cite{KTY}.
The double transverse-spin asymmetry 
for the  Drell-Yan (DY) process
in collisions of transversely polarized protons and antiprotons,
$p^{\uparrow}\bar{p}^{\uparrow}\rightarrow l^+l^-X$,
is given as ($d\omega \equiv dQ^2 dx_F d\Omega$)
\begin{equation}
A_{TT} =
\frac{d\sigma^{\uparrow \uparrow}/d\omega 
-d\sigma^{\uparrow\downarrow}/d\omega}
{d\sigma^{\uparrow \uparrow}/d\omega 
+d\sigma^{\uparrow\downarrow}/d\omega}              
       = \hat{a}_{TT}(\theta, \phi)  
\frac{\sum_{a} e_{a}^2
                   h_{1}^a(x_1,Q^2)h_{1}^{a}(x_2,Q^2)}
               {\sum_{a} e_{a}^2 f_{1}^a(x_1,Q^2)f_{1}^{a}(x_2,Q^2)},
\label{ATT}
\end{equation}
at the leading order (LO) QCD, for the dilepton production with the invariant mass $Q$ and 
the direction given by the angle
$\Omega= (\theta, \phi)$.
$h_1^a$ and $f_1^a$ denote the transversity and unpolarized quark-distributions
inside proton,
and the summation is over all quark and anti-quark flavors with $e_a$ the 
corresponding electric charge. 
The scaling variables $x_{1,2}$ 
represent the momentum fractions associated with the partons
annihilating via the DY mechanism, such that 
$Q^2
=(x_1 P_1 + x_2 P_2)^2 
= x_1 x_2 s$ and $x_F=x_1-x_2$,
where $s=(P_1+P_2)^2$ is
the CM energy squared of $p^{\uparrow}\bar{p}^{\uparrow}$.
$\hat{a}_{TT}(\theta, \phi)$
represents
the asymmetry in the parton level.
In particular, 
moderate energies at GSI, $30 \lesssim s \lesssim 200$~GeV$^2$,
allow us to measure (\ref{ATT}) for $0.2 \lesssim Q/\sqrt{s}\lesssim 0.7$,
and probe $h_1 (x_{1,2}, Q^2)$ in the ``valence region''.
It turns out~\cite{KKT:08} that
$A_{TT}$ 
at GSI is quite stable when including the QCD (resummation and fixed-order) corrections,
and that the recent empirical information on $h_1$~\cite{Anselmino:07} 
leads to 
large value of (\ref{ATT}) at GSI.
We note that the double longitudinal-spin asymmetry $A_{LL}$ 
for 
$p^{\rightarrow}\bar{p}^{\rightarrow}\rightarrow l^+l^-X$
is given by (\ref{ATT}) with the replacement
$h_1\rightarrow g_1$, $\hat{a}_{TT}\rightarrow \hat{a}_{LL}$,
where $g_1$ is the 
chiral-even, helicity quark-distribution and $\hat{a}_{LL}$
is the corresponding partonic asymmetry.

It should not be overlooked that 
the double-spin 
longitudinal-transverse asymmetry 
$A_{LT}=[d\sigma^{\rightarrow \uparrow}/d\omega
-d\sigma^{\rightarrow\downarrow}/d\omega]/
[ d\sigma^{\rightarrow \uparrow}d\omega
+d\sigma^{\rightarrow\downarrow}/d\omega]$ 
is also readily
accessible in the DY experiments
at GSI:
$A_{LT}$ plays a distinguished role in spin physics,
allowing us to access the twist-3 spin-dependent parton distributions 
as leading effects (see \cite{KTY}),
\begin{equation}
A_{LT}
       =
\frac{M}{Q} 
\hat{a}_{LT}(\theta, \phi)
\frac{\sum_{a} e_{a}^2
                   \left [g_{1}^a(x_1,Q^2)x_2g_{T}^{a}(x_2,Q^2)
                        + x_1h_{L}^a(x_1,Q^2)h_{1}^{a}(x_2,Q^2) \right]}
               {\sum_{a} e_{a}^2 f_{1}^a(x_1,Q^2)f_{1}^{a}(x_2,Q^2)},      
\label{ALT}
\end{equation}
with the proton mass $M$, although 
suppressed by $M/Q$ compared to $A_{TT}$, $A_{LL}$
that receive contribution 
only from twist-2 distributions.
$g_T^a$ and $h_L^a$ denote the 
twist-3 distributions
associated with the transversely and longitudinally polarized proton, respectively;
the chiral-even distribution 
$g_T^a$ is also accessible by the longitudinal-transverse asymmetry associated with the 
structure function $g_2$ in the polarized DIS~\cite{g2exp}, 
but the chiral-odd $h_L^a$
is not accessible by inclusive DIS.
We assess the potential of $A_{LT}$ at GSI experiments~\cite{KTY},
extending the previous study on $A_{LT}$ in $pp$-collision at RHIC energy~\cite{Kanazawa:1998rw}.

The distributions $h_L$ and $g_T$ 
contain the piece expressed by twist-2 matrix element 
as
\begin{equation}
h_L^a(x,Q^2) = 2x\int_{x}^{1}dy \frac{h_1^a(y,Q^2)}{y^2}  
+\cdots,
\;\;\;\;\;\;\;\;\;\;\;\;
g_T^a(x,Q^2) = \int_{x}^{1} dy\frac{g_1^a(y,Q^2)}{y}
+\cdots,
\label{hLWW}
\end{equation}
where the ellipses stand for ``genuine twist-3'' contributions 
given as matrix element of the twist-3 operators; those
twist-3 operators can be reexpressed as 
quark-gluon-quark three-body correlation operators on the lightcone,
using the QCD equations of motion (see \cite{KT,BBKT}).  
We call the twist-2 component, 
shown explicitly in (\ref{hLWW}), 
the ``Wandzura-Wilczek (WW)'' part. 
Because the operators with different ``geometric twist'' 
do not mix with each other under 
QCD evolution with $Q^2$,
both $x$- and $Q^2$-dependences of the WW part
are determined solely by those of the corresponding twist-2 distributions.

We work at LO QCD with (\ref{ALT}), which provides sufficient accuracy
for our first estimate of $A_{LT}$ at GSI. We may anticipate that
QCD corrections to $A_{LT}$ at GSI kinematics could be small, similarly to the case for $A_{TT}$
mentioned above.
We also employ the ``WW approximation''
to $A_{LT}$, i.e., evaluate (\ref{ALT}) using (\ref{hLWW}) with the genuine twist-3
contributions omitted: the WW approximation for $g_T$ of (\ref{hLWW}) 
is supported by
the data on the 
structure function $g_2$ from the polarized DIS experiments~\cite{g2exp}, as well as 
from lattice QCD simulation~\cite{RBRC}.
Estimates from nucleon models, combined with the QCD evolution which is different 
between the relevant twist-2~\cite{KMHKKV:97} and
twist-3 operators~\cite{KT,BBKT} such that the latter is more strongly suppressed for high $Q^2$
than the former, also suggest 
that the WW part of (\ref{hLWW}) 
dominates $h_L$ and $g_T$ for $Q^2 \gg 1$ GeV$^2$~\cite{Kanazawa:1998rw,KK}.

We use the LO GRV98~\cite{GRV:98} 
and GRSV2000 (``standard scenario'')~\cite{GRSV:00} 
distributions  
for the unpolarized and longitudinally-polarized quark distributions 
$f_1$ and $g_1$, respectively.
For the LO transversity distribution $h_1$,
we are guided by the recent information from the LO global fit~\cite{Anselmino:07}:
we find that 
a useful estimate can be obtained
by assuming the relation,
$h_1^a (x,\mu^2)=g_1^a(x,\mu^2)$,
at a low scale $\mu$ ($\mu^2=0.26$ GeV$^2$ using the GRSV2000 $g_1$);
its QCD evolution from $\mu^2$ to $Q^2$ is controlled by 
the LO DGLAP kernel~\cite{KMHKKV:97} for $h_1$.
Comparing the resulting transversity~\cite{KTY} 
with $h_1$ from the LO global 
fit~\cite{Anselmino:07},
our $h_1$ lies slightly outside the error band of the fit,
for the valence region $0.2 \lesssim x \lesssim 0.7$ relevant 
for the GSI kinematics.
Thus our $h_1$ will provide a realistic estimate of  
the upper bound of the relevant asymmetries using (\ref{ATT})-(\ref{hLWW}), 
implied by the present empirical uncertainty 
in transversity.
Similar choice for $h_1$ has been adopted 
to calculate QCD corrections for $A_{TT}$ 
at GSI~\cite{KKT:08}.
In the valence region relevant for the GSI kinematics,
the dominance of 
the $u$-quark contribution is observed for all twist-2 quark-distributions,
$(h_1^u)^2 \gg (h_1^a)^2$ with $a=\bar{u}, d, \bar{d}, \ldots$~\cite{KTY,Anselmino:07}, 
and similarly for $g_1$ and $f_1$~\cite{GRSV:00,GRV:98}.
Calculating (\ref{hLWW}) using our twist-2 distributions,
we find the similar $u$-quark dominance for $h_L$ and $g_T$ 
in the WW approximation~\cite{KTY}.
Therefore, from (\ref{ALT}) at GSI kinematics,
\begin{equation}
\widetilde{A}_{LT}\equiv 
\frac{A_{LT}}{(M/Q)\hat{a}_{LT}(\theta,\phi)} 
\simeq
\frac{g_{1}^u(x_1,Q^2)x_2\int_{x_2}^{1} dy\frac{g_1^u(y,Q^2)}{y}
 + 2x_1^2 \int_{x_1}^{1}dy \frac{h_1^u(y,Q^2)}{y^2}  h_{1}^{u}(x_2,Q^2)}
               {f_{1}^u(x_1,Q^2)f_{1}^u(x_2,Q^2)}
+ \cdots,
\label{ALT2}
\end{equation}
where the ellipses denote the 
contributions associated with the genuine twist-3 operators.
Thus the WW approximation to $A_{LT}$ at GSI is
directly related to 
the behavior of $h_1^u$.

\begin{figure}
\includegraphics[height=5.2cm]{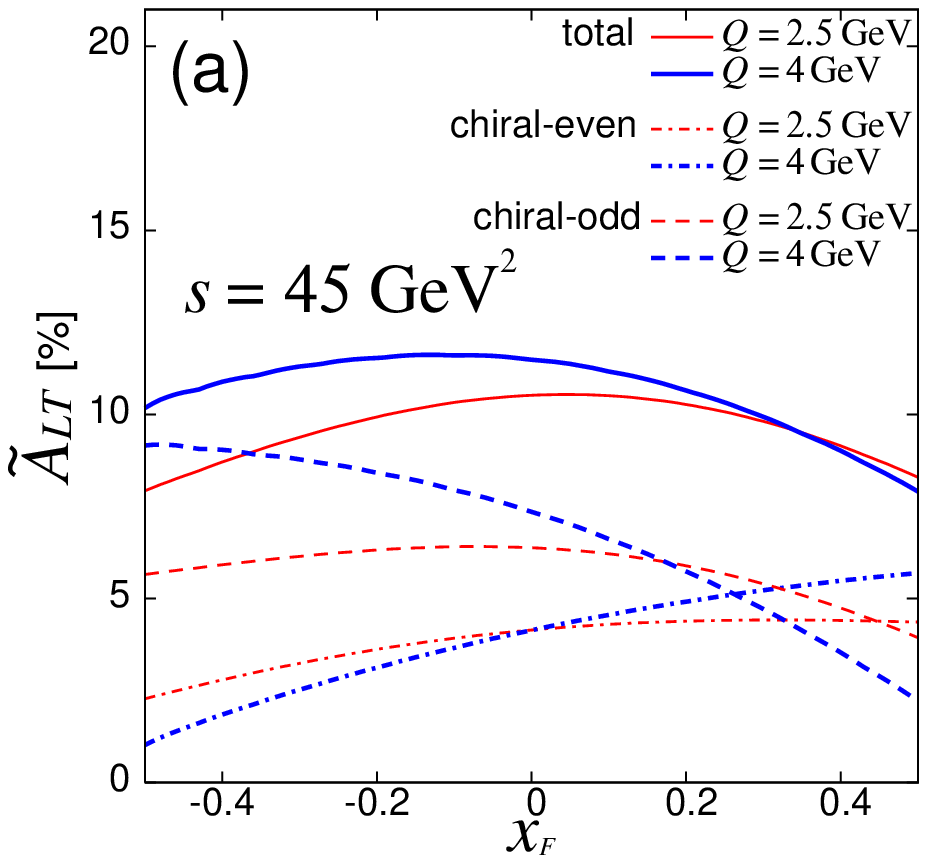}~~
\includegraphics[height=5.2cm]{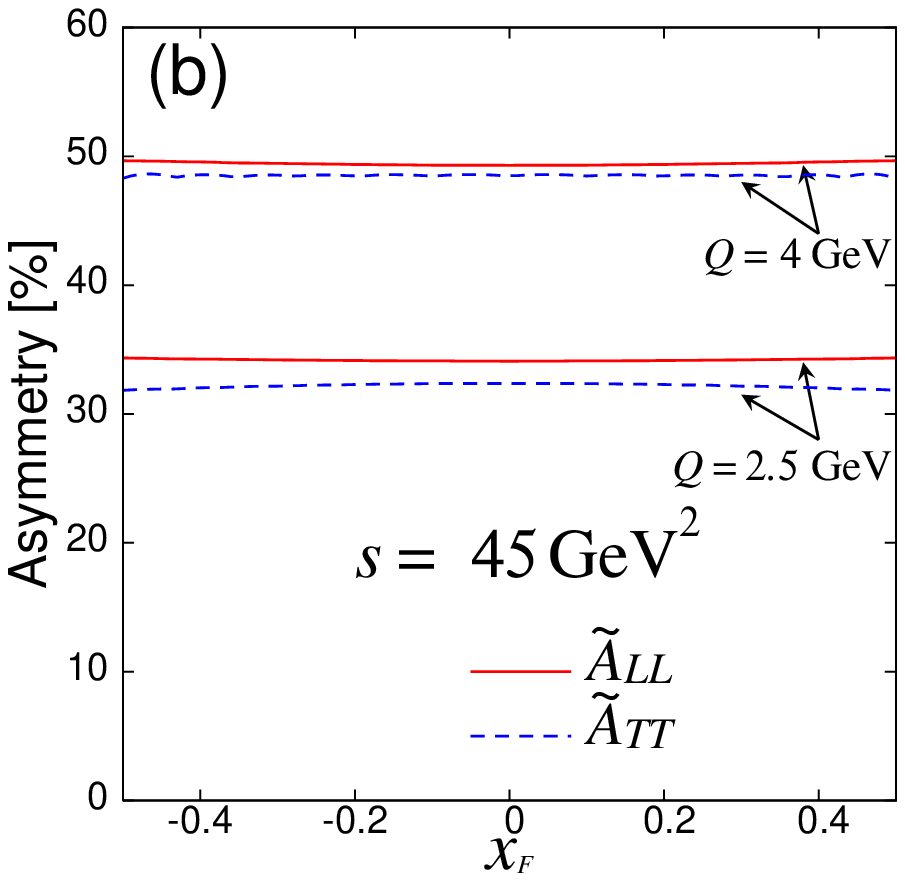}
\caption{(a)~$\widetilde{A}_{LT}$ and (b)~$\widetilde{A}_{LL}$ and $\widetilde{A}_{TT}$ 
as a function of $x_F$ for $Q=2.5$ and $4$ GeV at $s=45$ GeV$^2$.
}
\label{fig:2}
\end{figure}
We present the 
results~\cite{KTY} 
for the ``reduced asymmetries'',
$\widetilde{A}_{LT}$, defined in
(\ref{ALT2}),
and $\widetilde{A}_{YY}\equiv A_{YY}/\hat{a}_{YY}$
($Y= L, T$).
Figure \ref{fig:2}(a) shows $\widetilde{A}_{LT}$, along with
the separated contributions from the chiral-even and -odd
distributions corresponding to the first and second terms in the numerator in (\ref{ALT}).
The value of $\widetilde{A}_{LT}$ is significant.
Its $x_F$-behavior is in strong contrast to the flat and symmetric behavior of 
$\widetilde{A}_{LL}$, $\widetilde{A}_{TT}$ 
at the same kinematics, shown in Fig.~\ref{fig:2}(b).
$\widetilde{A}_{LT}$ is smaller than $\widetilde{A}_{TT}$, $\widetilde{A}_{LL}$  
by the additional $x_{1,2}$ factor
in (\ref{ALT}) compared with 
(\ref{ATT}), and also by
the behavior 
of $h_L$ and $g_T$ where
the integral for the WW part, 
with the factors $1/y^2$, $1/y$
in (\ref{hLWW}), shifts the peak of the distributions
to lower $x$ with the suppressed peak-height, compared with 
the corresponding twist-2 distributions~\cite{KTY}.
Indeed, for $pp$ collisions where the sea-quark region is probed,
these effects, in particular the additional $x_{1,2}$ factor, 
lead to $\widetilde{A}_{LT}$ much smaller than 
the corresponding 
$\widetilde{A}_{LL}$, $\widetilde{A}_{TT}$~\cite{Kanazawa:1998rw}.

\begin{figure}
\includegraphics[height=5.2cm]{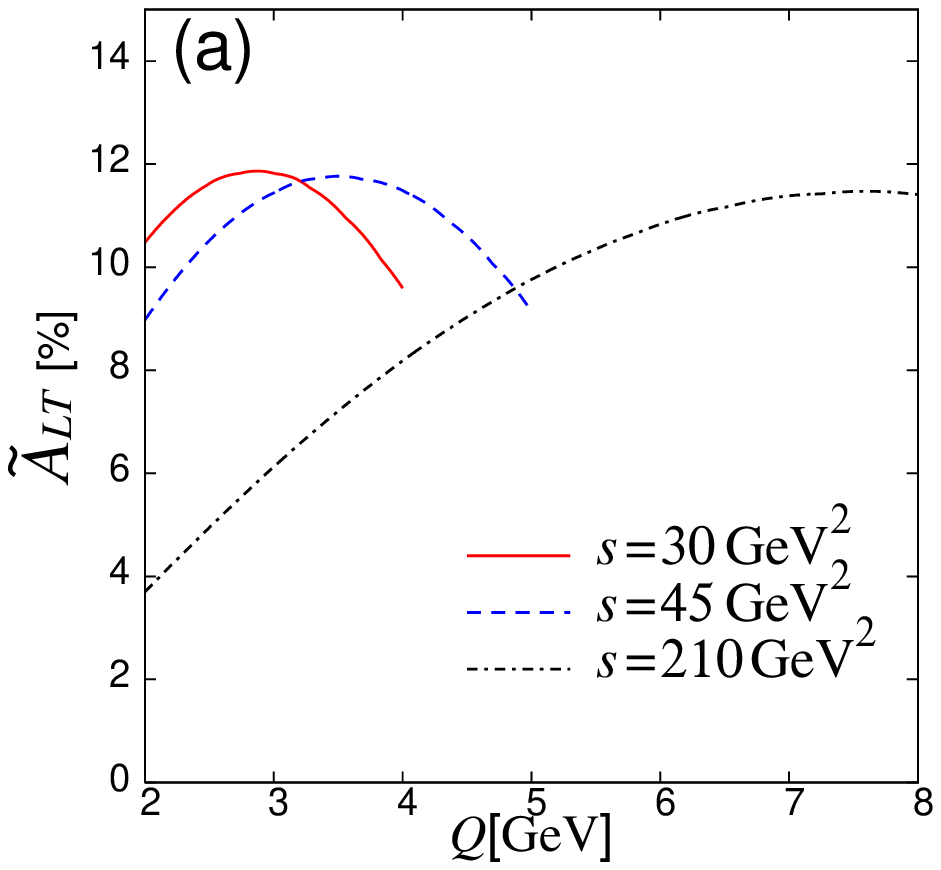}~~
\includegraphics[height=5.2cm]{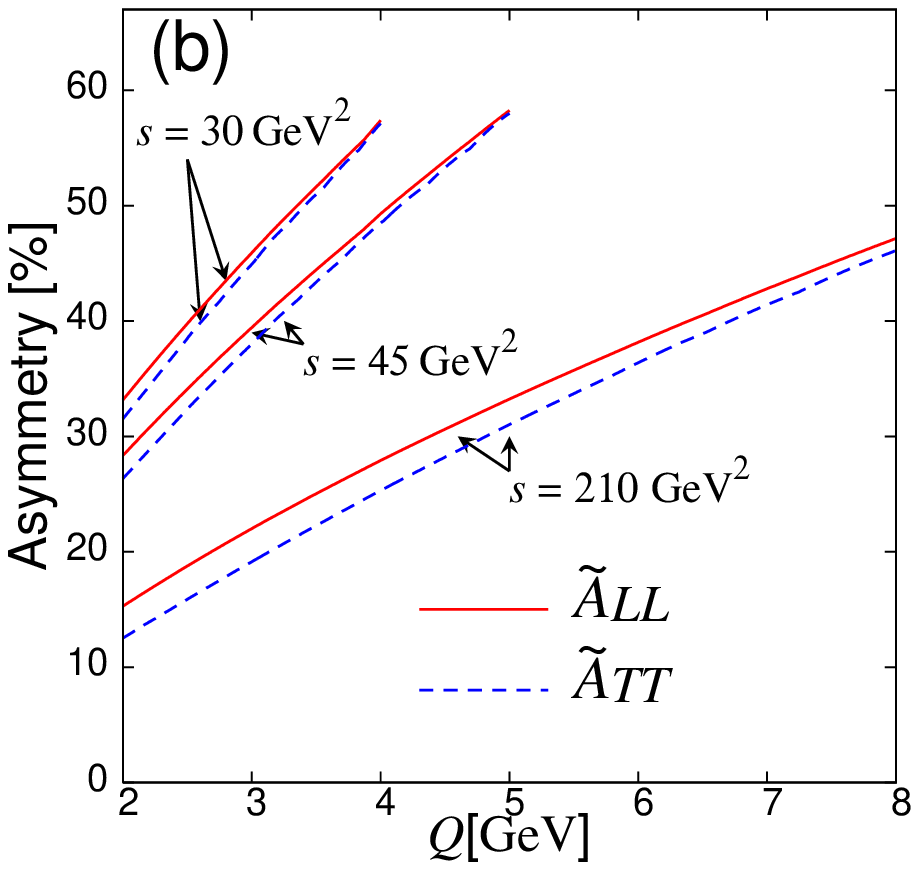}
\caption{(a)~$\widetilde{A}_{LT}$ and (b)~$\widetilde{A}_{LL}$ and $\widetilde{A}_{TT}$ 
at $x_F=0$ as a function of $Q$ 
for $s=30$, $45$ and $210$ GeV$^2$. 
}
\label{fig:4}
\end{figure}

Figure~\ref{fig:4} shows
the relevant asymmetries
with $x_1=x_2=Q/\sqrt{s}$ as a function of $Q$.
The $Q$ dependence of $\widetilde{A}_{TT}$ as well as $\widetilde{A}_{LL}$
directly reflects the $x$ dependence of the corresponding 
distributions, because the $u$-quark dominance in (\ref{ATT}) 
implies that $\widetilde{A}_{TT}$ is
controlled
by the ratio $h_{1}^u(x,Q^2)/f_{1}^u(x,Q^2)$ and 
the scale dependence 
in this ratio turns out to almost
cancel between the numerator and denominator 
in the valence region relevant at GSI~\cite{KKT:08}.
With GRV and GRSV parameterizations,
the ratio $h_{1}^u(x,1 {\rm GeV}^2)/f_{1}^u(x,1 {\rm GeV}^2)$, as well
as $g_{1}^u(x,1 {\rm GeV}^2)/f_{1}^u(x,1 {\rm GeV}^2)$,
is actually an increasing function 
of $x$, leading to the $Q$-dependence in Fig.~\ref{fig:4}(b) through $x=Q/\sqrt{s}$.
For $\widetilde{A}_{LT}$, however, the cancellation 
of the scale dependence
between the numerator and denominator in (\ref{ALT2}) is less complete
due to the additional 
$y$-integral for the WW part, which, combined with the additional
factor $x_1$ or $x_2$ ($=Q/\sqrt{s}$),
results in the novel $Q$-dependence in Fig.~\ref{fig:4}(a).

We have presented
a first realistic estimate of the WW contribution to $A_{LT}$ at GSI.
Our results indicate that rather large $A_{LT} \sim 10$\%  can be observed.
If the strong deviation from our results were observed 
in the future GSI measurements of $A_{LT}$, 
this would provide an indication of 
large twist-3 effect, associated with the quark-gluon-quark correlation inside nucleon, 
in particular, in the chiral-odd spin structure. 
Such data 
can be analyzed utilizing the QCD evolution of the three-body operators 
for large number of colors~\cite{BBKT},
which allows us to avoid the notorious complication in the evolution at higher twist.

\begin{theacknowledgments}
The work of Y.K. is supported 
by Uchida Energy Science Promotion Foundation.
The work of 
K.T. is supported by the Grant-in-Aid for Scientific Research 
No.~B-19340063. 
\end{theacknowledgments}


\end{document}